\documentclass[12pt,aps,nofootinbib,superscriptaddress]{revtex4}
\usepackage{bm}
\begin{document}
\title{Exclusive weak $\bm{B}$ decays to excited 
meson states}
\author{\firstname{D.} \surname{Ebert}}
\affiliation{Research Center for Nuclear Physics (RCNP), Osaka University,
Ibaraki, Osaka 567, Japan}
\affiliation{Institut f\"ur Physik, Humboldt--Universit\"at zu Berlin,
Invalidenstr.110, D-10115 Berlin, Germany}
\author{\firstname{R. N.} \surname{Faustov}}
\affiliation{Russian Academy of Sciences,
Scientific Council for Cybernetics,
Vavilov Street 40, Moscow 117333, Russia}
\author{\firstname{V. O.} \surname{Galkin}}      
\affiliation{Russian Academy of Sciences,
Scientific Council for Cybernetics,
Vavilov Street 40, Moscow 117333, Russia}
\begin{abstract}
{Exclusive semileptonic $B$ decays to 
radially excited charmed mesons and rare radiative $B$ decay to the
orbitally excited tensor $K_2^*(1430)$ meson 
are investigated in the framework of the  relativistic 
quark model based on the quasipotential approach in quantum field theory.
The heavy quark expansion is applied for the semileptonic $B\to D^{(*)}{}'e\nu$
decays. Both relativistic and the $1/m_Q$ corrections are found to play 
an important role for these decays and substantially
modify results. For rare radiative $B\to K_2^*(1430)\gamma$ decay 
such an expansion is not used for the $s$ quark. 
Instead we apply the expansion 
in inverse powers of large recoil momentum of final $K_2^*(1430)$ meson.
The calculated branching fraction 
$BR(B\to K_2^*(1430)\gamma)=(1.7\pm 0.6)\times
10^{-5}$ as well as the ratio $BR(B\to K_2^*(1430)\gamma)/BR(B\to
K^*(892)\gamma)=0.38\pm 0.08$ are found to be in a good agreement with
recent experimental data from CLEO. }
\end{abstract}

\maketitle 

\section{Introduction}
The investigation of weak $B$ decays to excited
mesons presents a problem interesting both from the experimental and
theoretical point of view. The current experimental data on semileptonic
$B$ decays to ground state $D$ mesons indicate that a substantial part  
($\approx 40\%$) of the inclusive semileptonic $B$ decays should go to 
excited $D$ meson states. First experimental data on some 
exclusive $B$ decay channels to excited charmed mesons are becoming 
available now \cite{cleo,aleph,opal} and more data are expected in near
future. Thus the comprehensive theoretical study of these decays is 
necessary. The presence of a heavy 
quark in the initial and final meson states in these decays considerably
simplifies their theoretical description. A good starting point for this
analysis is the infinitely heavy quark limit, $m_Q\to\infty$ \cite{iw}. 
In this limit the heavy quark symmetry arises, which strongly
reduces the number of independent weak form factors \cite{iw1}.
The heavy quark mass and spin then decouple and all meson properties are
determined by light-quark degrees of freedom alone. This leads to
a considerable reduction of the number of independent form factors
which are necessary for the description of heavy-to-heavy semileptonic
decays. For example, in this limit only one form factor is necessary 
for the semileptonic
$B$ decay to $S$-wave $D$ mesons (both for the ground state and its
radial excitations), while the decays to $P$ states require two form
factors \cite{iw1}. It is important to note 
that in the infinitely heavy quark limit matrix elements between
a $B$ meson and an excited $D$ meson should vanish at the point of
zero recoil of the final excited charmed meson in the rest frame of the $B$
meson. In the case of $B$ decays to radially excited charmed mesons
this follows from the orthogonality of radial parts of wave functions,
while for the decays to orbital excitations this is the consequence of
orthogonality of their angular parts. However, some of the $1/m_Q$ corrections
to these decay matrix elements give nonzero contributions at zero 
recoil. As a result the role of these  corrections could be considerably
enhanced, since the kinematical range for $B$ decays to excited states is
a rather small region around zero recoil. 

Rare radiative decays of $B$ mesons are induced by flavour
changing neutral currents and thus they can serve as sensitive probes 
of new physics beyond the standard model.
Such decays are governed by one-loop (penguin) diagrams with
the main contribution from a virtual top quark and a $W$ boson. The
statistics of rare radiative $B$ decays considerably increased since
the first observation of the $B\to K^*\gamma$ decay in 1993 by  
CLEO \cite{cleo1}.
This allowed a significantly more precise determination of exclusive and 
inclusive branching fractions \cite{cleo2}. Recently the first observation 
of rare $B$ decay to the orbitally excited tensor strange meson
$B\to K_2^*(1430)\gamma$ has been reported by
CLEO \cite{cleo2} with a branching fraction
\begin{equation}\label{br}
BR(B\to K_2^*(1430)\gamma)=(1.66^{+0.59}_{-0.53}\pm 0.13)\times 10^{-5},
\end{equation}
as well as the ratio of exclusive branching fractions
\begin{equation}\label{ratio}
r\equiv
\frac{BR(B\to K_2^*(1430)\gamma )}{BR(B\to K^*(892)\gamma)}
=0.39^{+0.15}_{-0.13}.
\end{equation}
These new experimental data provide a challenge to the theory.
Many theoretical approaches have been employed to predict the
exclusive $B\to K^*(892)\gamma$ decay rate (for a review see
\cite{lss} and references therein). Less attention has been payed
to rare radiative $B$ decays to excited strange mesons
\cite{a,aom,vo}. Most of these theoretical approaches
\cite{aom,vo}  rely on the heavy quark limit both for the initial
$b$ and final $s$ quark and the nonrelativistic quark model.
However, the two predictions \cite{aom,vo} for the ratio $r$ in
Eq.~(\ref{ratio}) differ by an order of magnitude, due to a
different treatment of the long distance effects and, as a
result, a different determination of corresponding Isgur-Wise
functions. Only the prediction of Ref.~\cite{vo} is consistent
with data (\ref{br}), (\ref{ratio}). Nevertheless, it is
necessary to point out that the $s$ quark in the final $K^*$
meson is not heavy enough, compared to the $\bar \Lambda$
parameter, which determines the scale of $1/m_Q$ corrections in
heavy quark effective theory \cite{n}. Thus the $1/m_s$ expansion
is not appropriate. Notwithstanding, the ideas of heavy quark
expansion can be applied to the exclusive $B\to
K^*(K_2^*)\gamma$  decays. From the kinematical analysis it
follows that the final $K^*(K_2^*)$ meson bears a large
relativistic recoil momentum $\vert {\bm\Delta} \vert$ of order
of $m_b/2$ and an energy of the same order. So it is possible to
expand the matrix element of the effective Hamiltonian both in
inverse powers of the $b$ quark mass for the initial state and in
inverse powers of the recoil momentum $\vert{\bm\Delta} \vert$
for the final state \cite{gf,cbf}. 

Our relativistic quark model is  based on the quasipotential approach in
quantum field
theory with a specific choice  of the quark-antiquark interaction
potential. It provides
a consistent scheme for the  calculation of all relativistic corrections
at a given $v^2/c^2$ 
order and allows for the heavy  quark $1/m_Q$ expansion. In preceding
papers 
we applied this model to the calculation of the mass spectra of 
orbitally and radially excited states of heavy-light mesons \cite{egf},
as well as to the description of the rare radiative decay $B\to K^*\gamma$
\cite{gf} and of weak decays of $B$ mesons to ground state
heavy and light mesons \cite{fgm,efg}. The heavy quark expansion for the ground
state heavy-to-heavy semileptonic transitions \cite{fg} was found to be 
in agreement with model-independent predictions of the heavy quark effective 
theory (HQET).

\section{Relativistic quark model}
In our model a meson is described by the wave
function of the bound quark-antiquark state, which satisfies the
quasipotential equation \cite{3} of the Schr\"odinger type~\cite{4}:
\begin{equation}
\label{quas}
{\left(\frac{b^2(M)}{2\mu_{R}}-\frac{{\bf
p}^2}{2\mu_{R}}\right)\Psi_{M}({\bf p})} =\int\frac{d^3 q}{(2\pi)^3}
 V({\bf p,q};M)\Psi_{M}({\bf q}),
\end{equation}
where the relativistic reduced mass is
\begin{equation}
\mu_{R}=\frac{M^4-(m^2_q-m^2_Q)^2}{4M^3}.
\end{equation}
In the center-of-mass system the relative momentum squared on mass shell
reads
\begin{equation}
{b^2(M) }
=\frac{[M^2-(m_q+m_Q)^2][M^2-(m_q-m_Q)^2]}{4M^2}.
\end{equation}

The kernel
$V({\bf p,q};M)$ in Eq.~(\ref{quas}) is the quasipotential operator of
the quark-antiquark interaction. It is constructed with the help of the
off-mass-shell scattering amplitude, projected onto the positive
energy states. An important role in this construction is played
by the Lorentz-structure of the confining quark-antiquark interaction
in the meson.  In
constructing the quasipotential of the quark-antiquark interaction
we have assumed that the effective
interaction is the sum of the usual one-gluon exchange term and the mixture
of vector and scalar linear confining potentials.
The quasipotential is then defined by
\cite{mass}
\begin{eqnarray}
\label{qpot}
\!\!\!\!\!\! V({\bf p,q};M)&=&\bar{u}_q(p)\bar{u}_Q(-p){\cal V}({\bf p}, {\bf
q};M)u_q(q)u_Q(-q)\cr
&=&\bar{u}_q(p)
\bar{u}_Q(-p)\Bigg\{\frac{4}{3}\alpha_sD_{ \mu\nu}({\bf
k})\gamma_q^{\mu}\gamma_Q^{\nu}
+V^V_{\rm conf}({\bf k})\Gamma_q^{\mu}
\Gamma_{Q;\mu}+V^S_{\rm conf}({\bf
k})\Bigg\}u_q(q)u_Q(-q),
\end{eqnarray}
where $\alpha_s$ is the QCD coupling constant, $D_{\mu\nu}$ is the
gluon propagator in the Coulomb gauge
and ${\bf k=p-q}$; $\gamma_{\mu}$ and $u(p)$ are
the Dirac matrices and spinors
\begin{equation}
\label{spinor}
u^\lambda({p})=\sqrt{\frac{\epsilon(p)+m}{2\epsilon(p)}}
\left(\begin{array}{c}
1\\ \displaystyle\frac{\mathstrut{\bm{\sigma }{\bf p}}}{\mathstrut\epsilon(p)+m}
\end{array}\right)
\chi^\lambda
\end{equation}
with $\epsilon(p)=\sqrt{{\bf p}^2+m^2}$.
The effective long-range vector vertex is
given by
\begin{equation}
\Gamma_{\mu}({\bf k})=\gamma_{\mu}+
\frac{i\kappa}{2m}\sigma_{\mu\nu}k^{\nu},
\end{equation}
where $\kappa$ is the Pauli interaction constant characterizing the
nonperturbative anomalous chromomagnetic moment of quarks. Vector and
scalar confining potentials in the nonrelativistic limit reduce to
\begin{equation}\label{vconf}
V^V_{\rm conf}(r)=(1-\varepsilon)(Ar+B),\qquad
V^S_{\rm conf}(r) =\varepsilon (Ar+B),
\end{equation}
reproducing
\begin{equation}
V_{\rm conf}(r)=V^S_{\rm conf}(r)+
V^V_{\rm conf}(r)=Ar+B,
\end{equation}
where $\varepsilon$ is the mixing coefficient.

The quasipotential for the heavy quarkonia,
expanded in $v^2/c^2$, can be found in Refs.~\cite{mass,pot} and for
heavy-light mesons in \cite{egf}.
All the parameters of
our model, such as quark masses, parameters of the linear confining potential,
mixing coefficient $\varepsilon$ and anomalous
chromomagnetic quark moment $\kappa$, were fixed from the analysis of
heavy quarkonia masses \cite{mass} and radiative decays \cite{gfm}.
The quark masses
$m_b=4.88$ GeV, $m_c=1.55$ GeV, $m_s=0.50$ GeV, $m_{u,d}=0.33$ GeV and
the parameters of the linear potential $A=0.18$ GeV$^2$ and $B=-0.30$ GeV
have the usual quark model values.
In Ref.~\cite{fg} we have considered the expansion of  the matrix
elements of weak heavy quark currents between pseudoscalar and vector
meson ground states up to the second order in inverse powers of
the heavy quark
masses. It has been found that the general structure of the leading,
first,
and second order $1/m_Q$ corrections in our relativistic model is in accord
with the predictions of HQET. The heavy quark symmetry and QCD impose rigid
constraints on the parameters of the long-range potential in our model.
The analysis
of the first order corrections \cite{fg} fixes the value of the
Pauli interaction
constant $\kappa=-1$. The same value of $\kappa$  was found previously
from  the fine splitting of heavy quarkonia ${}^3P_J$- states \cite{mass}.
The value of the parameter mixing
vector and scalar confining potentials $\varepsilon=-1$
was found from the analysis of the second order corrections \cite{fg}.
This value is very close to the one determined from considering radiative
decays of heavy quarkonia \cite{gfm}.

\begin{figure}
\unitlength=0.9mm
\begin{picture}(150,150)
\put(10,100){\line(1,0){50}}
\put(10,120){\line(1,0){50}}
\put(35,120){\circle*{5}}
\multiput(32.5,130)(0,-10){2}{\begin{picture}(5,10)
\put(2.5,10){\oval(5,5)[r]}
\put(2.5,5){\oval(5,5)[l]}\end{picture}}
\put(5,120){$b$}
\put(5,100){$\bar q$}
\put(5,110){$B$}
\put(65,120){$c,s$}
\put(65,100){$\bar q$}
\put(65,110){$\!\!\!\!\!\!\!\!\!D^{(*)}{}', K_2^*$}
\put(43,140){$W,\gamma$}
\put(0,85){\small FIG. 1. Lowest order vertex function $\Gamma^{(1)}$
corresponding to Eq.~(\ref{gamma1}). }
\put(10,20){\line(1,0){50}}
\put(10,40){\line(1,0){50}}
\put(25,40){\circle*{5}}
\put(25,40){\thicklines \line(1,0){20}}
\multiput(25,40.5)(0,-0.1){10}{\thicklines \line(1,0){20}}
\put(25,39.5){\thicklines \line(1,0){20}}
\put(45,40){\circle*{1}}
\put(45,20){\circle*{1}}
\multiput(45,40)(0,-4){5}{\line(0,-1){2}}
\multiput(22.5,50)(0,-10){2}{\begin{picture}(5,10)
\put(2.5,10){\oval(5,5)[r]}
\put(2.5,5){\oval(5,5)[l]}\end{picture}}
\put(5,40){$b$}
\put(5,20){$\bar q$}
\put(5,30){$B$}
\put(65,40){$c,s$}
\put(65,20){$\bar q$}
\put(65,30){$\!\!\!\!\!\!\!\!\!D^{(*)}{}', K_2^*$}
\put(33,60){$W,\gamma$}
\put(90,20){\line(1,0){50}}
\put(90,40){\line(1,0){50}}
\put(125,40){\circle*{5}}
\put(105,40){\thicklines \line(1,0){20}}
\multiput(105,40.5)(0,-0.1){10}{\thicklines \line(1,0){20}}
\put(105,39,5){\thicklines \line(1,0){20}}
\put(105,40){\circle*{1}}
\put(105,20){\circle*{1}}
\multiput(105,40)(0,-4){5}{\line(0,-1){2}}
\multiput(122.5,50)(0,-10){2}{\begin{picture}(5,10)
\put(2.5,10){\oval(5,5)[r]}
\put(2.5,5){\oval(5,5)[l]}\end{picture}}
\put(85,40){$b$}
\put(85,20){$\bar q$}
\put(85,30){$B$}
\put(145,40){$c,s$}
\put(145,20){$\bar q$}
\put(145,30){$\!\!\!\!\!\!\!\!\!D^{(*)}{}', K_2^*$}
\put(133,60){$W,\gamma$}
\put(0,5){\makebox[14cm][s]{\small FIG. 2. Vertex function $\Gamma^{(2)}$
corresponding to Eq.~(\ref{gamma2}). Dashed lines represent the}}
\put(0,0) {\makebox[14cm][s]{\small  effective   potential ${\cal V}$ in
Eq.~(\ref{qpot}). Bold lines denote the negative-energy part of the} }
\put(0,-5){\small quark propagator. }

\end{picture}
\bigskip
\end{figure}

In the quasipotential approach,  the matrix element of the weak
current $J_\mu=\bar f G b$ ($f=\{c$ or $s\}$, $G=\{\gamma^\mu(1-\gamma^5)$ or  
$\frac{i}{2} k^\nu \sigma_{\mu\nu}(1+\gamma^5)\}$) 
between the states of a $B$ meson
and an excited $F$ ($D^{(*)}{}'$ or $K_2^*$) meson has the form \cite{f}
\begin{equation}\label{mxet}
\langle F \vert J_\mu (0) \vert B\rangle
=\int \frac{d^3p\, d^3q}{(2\pi )^6} \bar \Psi_{F}({\bf
p})\Gamma _\mu ({\bf p},{\bf q})\Psi_B({\bf q}),\end{equation}
where $\Gamma _\mu ({\bf p},{\bf
q})$ is the two-particle vertex function and  $\Psi_{B,F}$ are the
meson wave functions projected onto the positive energy states of
quarks and boosted to the moving reference frame.
 The contributions to $\Gamma$ come from Figs.~1 and 2.
The contribution $\Gamma^{(2)}$ is the consequence of the
projection onto the positive-energy states. Note that the form
of the relativistic corrections resulting from the vertex function
$\Gamma^{(2)}$ explicitly depends on the Lorentz structure of the
$q\bar q$-interaction.  The vertex functions look like
\begin{equation}\label{gamma1}
\Gamma_\mu ^{(1)}({\bf p},{\bf q})=\bar
u_f(p_1) G u_b(q_1)(2\pi)^3\delta({\bf p}_2-{\bf q}_2),\end{equation}
and
\begin{eqnarray}\label{gamma2}
\Gamma_\mu^{(2)}({\bf p},{\bf q})&=&\bar u_f(p_1)\bar
u_q(p_2) \biggl\{ G \frac{\Lambda_b^{(-)}({ k}_1)}{ \epsilon
_b(k_1)+\epsilon_b(p_1)}\gamma_1^0{\cal V}({\bf p}_2-{\bf q}_2)\nonumber\\
& & +{\cal V}({\bf p}_2-{\bf q}_2)\frac{\Lambda_f^{(-)}(k_1')}{
\epsilon_f(k_1')+ \epsilon_f(q_1)}\gamma_1^0 G \biggr\}u_b(q_1) u_q(q_2), 
\end{eqnarray}
where ${\bf k}_1={\bf p}_1-{\bm\Delta};\quad {\bf k}_1'={\bf
q}_1+{\bm\Delta};\quad {\bm\Delta}={\bf p}_{F}-{\bf p}_B$;
$$\Lambda^{(-)}(p)=\frac{\epsilon(p)-\bigl( m\gamma ^0+\gamma^0({\bm{
\gamma} \bf p})\bigr) }{2\epsilon (p)}.$$

It is important to note that the wave functions entering the weak current
matrix element (\ref{mxet}) cannot  be both in the rest frame.
In the $B$ meson rest frame, the $F$ meson is moving with the recoil
momentum ${\bm \Delta}$. The wave function
of the moving $F$ meson $\Psi_{F\,{\bm\Delta}}$ is connected
with the $F$ wave function in the rest frame
$\Psi_{F\,{\bf 0}}\equiv \Psi_{F}$ by the transformation \cite{f}
\begin{equation}
\label{wig}
\Psi_{F\,{\bm\Delta}}({\bf
p})=D_f^{1/2}(R_{L_{\bm\Delta}}^W)D_q^{1/2}(R_{L_{
\bm\Delta}}^W)\Psi_{F\,{\bf 0}}({\bf p}),
\end{equation}
where $R^W$ is the Wigner rotation, $L_{\bm\Delta}$ is the Lorentz boost
from the meson rest frame to a moving one. The wave functions of $B$,
$D^{(*)}{}'$ and
$K_2^*$ mesons at rest were calculated by numerical solution of the
quasipotential equation (\ref{quas}).

\section{Semileptonic decays to radially excited states}

The matrix elements of the vector  and axial vector currents  between the
$B$ and radially excited $D'$ or $D^{*}{'}$ mesons can be parameterized by six
hadronic form factors:
 \begin{eqnarray}\label{ff}
 \frac{\langle D'(v')| \bar c\gamma^\mu b |B(v)\rangle  
}{\sqrt{m_{D'}m_B}}
 & =& h_+ (v+v')^\mu + h_- (v-v')^\mu , \cr
\langle D'(v')| \bar c\gamma^\mu  \gamma_5 b|B(v)\rangle 
 & =& 0, \cr
\frac{\langle D^*{'}(v',\epsilon)| \bar c\gamma^\mu b |B(v)\rangle  
}{\sqrt{m_{D^*{'}}m_B}}&
  =& i h_V \varepsilon^{\mu\alpha\beta\gamma} 
  \epsilon^*_\alpha v'_\beta v_\gamma ,\cr
\frac{\langle D^*{'}(v',\epsilon)| \bar c\gamma^\mu\gamma_5 b |B(v)\rangle  
}{\sqrt{m_{D^*{'}}m_B}}
& = &h_{A_1}(w+1) \epsilon^{* \mu} 
-(h_{A_2} v^\mu + h_{A_3} v'{}^\mu) (\epsilon^*\cdot v) ,
   \end{eqnarray}
where $v~(v')$ is the four-velocity of the $B~(D^{(*)}{'})$ meson,
$\epsilon^\mu$ is a  polarization vector  of the final vector
charmed meson, and the form factors $h_i$  are dimensionless 
functions of the product of velocities $w=v\cdot v'$.    

The HQET analysis \cite{rad} shows that five independent functions 
$\tilde\xi_3$, $\chi_b$ and
$\tilde\chi_{1,2,3}$, as well as two mass parameters
$\bar\Lambda$ and $\bar\Lambda^{(n)}$ are necessary 
to describe first order $1/m_Q$
corrections to  matrix elements of $B$ meson decays to radially excited
$D$ meson states. The function $\tilde\xi_3$ emerges from corrections to
the current in effective theory, while $\chi_b$ and
$\tilde\chi_{1,2,3}$ parameterize corrections to HQET Lagrangian. 
The resulting structure of the decay form factors is \cite{rad}
\begin{eqnarray}\label{cff}
h_{+}&=&\xi^{(n)}+\varepsilon_c\left[2\tilde\chi_1-4(w-1)\tilde\chi_2+
12\tilde\chi_3\right]+\varepsilon_b\chi_b,\cr
h_{-}&=&\varepsilon_c\left[2\tilde\xi_3-\left(\bar\Lambda^{(n)}+
\frac{\bar\Lambda^{(n)}-\bar\Lambda}{w-1}\right)\xi^{(n)}\right]
- \varepsilon_b\left[2\tilde\xi_3-\left(\bar\Lambda-
\frac{\bar\Lambda^{(n)}-\bar\Lambda}{w-1}\right)\xi^{(n)}\right],\cr
h_V&=&\xi^{(n)}+\varepsilon_c\biggl[2\tilde\chi_1+
\left(\bar\Lambda^{(n)}+\frac{\bar\Lambda^{(n)}
-\bar\Lambda}{w-1}\right)\xi^{(n)}
-4\tilde\chi_3\biggr]+\varepsilon_b\left[\chi_b+
\left(\bar\Lambda-\frac{\bar\Lambda^{(n)}-\bar\Lambda}{w-1}\right)\xi^{(n)}
-2\tilde\xi_3\right],\cr
h_{A_1}&=&\xi^{(n)}+\varepsilon_c\biggl[2\tilde\chi_1-4\tilde\chi_3
+\frac{w-1}{w+1}\left(\bar\Lambda^{(n)}+\frac{\bar\Lambda^{(n)}
-\bar\Lambda}{w-1}\biggr)\xi^{(n)}\right]\cr
&&+\varepsilon_b\left\{\chi_b+
\frac{w-1}{w+1}\left[
\left(\bar\Lambda-\frac{\bar\Lambda^{(n)}-\bar\Lambda}{w-1}\right)\xi^{(n)}
-2\tilde\xi_3\right]\right\},\cr
h_{A_2}&=&\varepsilon_c\biggl\{4\tilde\chi_2-\frac2{w+1}\biggl[
\left(\bar\Lambda^{(n)}+\frac{\bar\Lambda^{(n)}
-\bar\Lambda}{w-1}\right)\xi^{(n)}
+\tilde\xi_3\biggr]\biggr\},\cr
h_{A_3}&=&\xi^{(n)}+\varepsilon_c\biggl[2\tilde\chi_1-4\tilde\chi_2-
4\tilde\chi_3
+\frac{w-1}{w+1}\left(\bar\Lambda^{(n)}+\frac{\bar\Lambda^{(n)}
-\bar\Lambda}{w-1}\right)\xi^{(n)}-\frac2{w+1}\tilde\xi_3\biggr]\cr
&&+\varepsilon_b\left[\chi_b+
\left(\bar\Lambda-\frac{\bar\Lambda^{(n)}-\bar\Lambda}{w-1}\right)\xi^{(n)}
-2\tilde\xi_3\right],
\end{eqnarray}
where $\varepsilon_Q=1/(2m_Q)$ and $\bar\Lambda(\bar\Lambda^{(n)})=M(M^n)-m_Q$.

Now we can perform the heavy quark expansion for the matrix elements
of $B$ decays to excited $D$ mesons in the framework of our model and
determine leading and subleading Isgur--Wise functions. To do this we 
substitute the vertex functions $\Gamma^{(1)}$  and $\Gamma^{(2)}$ 
given by Eqs.~(\ref{gamma1}) and (\ref{gamma2})
in the decay matrix element (\ref{mxet})  and take into account the wave
function properties (\ref{wig}).
The resulting structure of this matrix element is
rather complicated, because it is necessary  to integrate both over  $d^3 p$
and $d^3 q$. The $\delta$ function in expression  (\ref{gamma1}) permits
us to perform one of these integrations and thus this contribution  can be 
easily calculated. The calculation  
of the vertex function $\Gamma^{(2)}$ contribution is  more difficult.
Here, instead
of a $\delta$ function, we have a complicated structure, containing the 
$Q\bar q$ interaction potential in the meson. 
However, we can expand this contribution in inverse
powers of heavy ($b,c$) quark masses and then use the  quasipotential
equation in
order to perform one of the integrations in the current matrix element.  
We carry out  the heavy quark expansion up to first order in $1/m_Q$. 
It is easy to see  that the vertex
function $\Gamma^{(2)}$ contributes already at  the subleading order of  
the $1/m_Q$ expansion. Then we compare the arising  decay matrix elements with
the form factor decompositions (\ref{cff}) for decays to radial excitations 
and determine the form factors. We find that, for the chosen values of our 
model parameters (the mixing
coefficient of vector and scalar confining potential $\varepsilon=-1$  and
the Pauli constant $\kappa=-1$), the resulting structure  at leading
and subleading order in $1/m_Q$ coincides with the model-independent  
predictions of HQET. This allows us to determine leading and subleading 
Isgur-Wise functions \cite{rad}:
\begin{eqnarray}
\label{xi}  
\xi^{(1)}(w)&=&\left(\frac{2}{w+1}\right)^{1/2}\!\!\!\!\int\!\!\frac{d^3 p}
{(2\pi)^3}\bar\psi^{(0)}_{D^{(*)}{'}}\!\!\left({\bf p}+
\frac{2\epsilon_q}{M_{D^{(*)}{'}}(w+1)}
{\bm \Delta}\right)
\psi^{(0)}_B({\bf p}),\\
\label{xi3}
\tilde\xi_3(w)&=&\left(\frac{\bar\Lambda^{(1)}+\bar\Lambda}2-m_q+
\frac16\frac{\bar\Lambda^{(1)}-\bar\Lambda}{w-1}\right)
\left(1+
\frac23\frac{w-1}{w+1}\right)\xi^{(1)}(w),\\ 
\label{chi1}
\tilde\chi_1(w)&\cong& \frac1{20}\frac{w-1}{w+1}\frac{\bar\Lambda^{(1)}
-\bar\Lambda}{w-1}\xi^{(1)}(w)\cr 
&&+\frac{\bar\Lambda^{(1)}}2
\left(\frac{2}{w+1}\right)^{1/2}\!\!\!\!
\int\!\!\frac{d^3 p}
{(2\pi)^3}\bar\psi^{(1)si}_{D^{(*)}{'}}\!\!\left({\bf p}+
\frac{2\epsilon_q}{M_{D^{(*)}{'}}(w+1)}
{\bm \Delta}\right)
\psi^{(0)}_B({\bf p}),\\
\label{chi2}
\tilde\chi_2(w)&\cong& -\frac1{12}\frac{1}{w+1}\frac{\bar\Lambda^{(1)}
-\bar\Lambda}{w-1}\xi^{(1)}(w),\\ 
\label{chi3}
\tilde\chi_3(w)&\cong& -\frac3{80}\frac{w-1}{w+1}\frac{\bar\Lambda^{(1)}
-\bar\Lambda}{w-1}\xi^{(1)}(w)\cr 
&&+\frac{\bar\Lambda^{(1)}}4
\left(\frac{2}{w+1}\right)^{1/2}\!\!\!\!
\int\!\!\frac{d^3 p}
{(2\pi)^3}\bar\psi^{(1)sd}_{D^{(*)}{'}}\!\!\left({\bf p}+
\frac{2\epsilon_q}{M_{D^{(*)}{'}}(w+1)}
{\bm \Delta}\right)
\psi^{(0)}_B({\bf p}),\\
\label{chib}
\chi_b(w)&\cong& \bar\Lambda\left(\frac{2}{w+1}\right)^{1/2}\!\!\!\!
\int\!\!\frac{d^3 p}
{(2\pi)^3}\bar\psi^{(0)}_{D^{(*)}{'}}\!\!\biggl({\bf p}+
\frac{2\epsilon_q}{M_{D^{(*)}{'}}(w+1)}
{\bm \Delta}\biggr)
\left[\psi^{(1)si}_B({\bf p})-3\psi^{(1)sd}_B({\bf p})\right]\!\!,
\end{eqnarray}
where ${\bm \Delta}^2=M_{D^{(*)}{'}}^2(w^2-1)$.
Here we used the expansion for the $S$-wave meson wave function
$$\psi_M=\psi_M^{(0)}+\bar\Lambda_M\varepsilon_Q\left(\psi_M^{(1)si}
+d_M\psi_M^{(1)sd}\right)+\cdots,$$
where $\psi_M^{(0)}$ is the wave function in the limit $m_Q\to\infty$,
$\psi_M^{(1)si}$ and $\psi_M^{(1)sd}$ are the spin-independent and 
spin-dependent first order $1/m_Q$ corrections, $d_P=-3$ for pseudoscalar and
$d_V=1$ for vector mesons.
The symbol $\cong$ in the expressions (\ref{chi1})--(\ref{chib}) for the
subleading functions $\tilde\chi_i(w)$ means that the corrections
suppressed by an additional power of the ratio $(w-1)/(w+1)$, which is equal 
to zero  at $w=1$ and less than $1/6$ at $w_{\rm max}$, were neglected. 
Since the main contribution to the decay rate comes from the values of 
form factors  close to $w=1$, these corrections turn out to be unimportant. 
 
It is clear from the expression (\ref{xi}) that the leading order contribution
vanishes at the point of zero recoil (${\bm \Delta}=0, w=1$) of the 
final $D^{(*)}{'}$ meson, 
since the radial parts of the wave functions $\Psi_{D^{(*)}{'}}$ and $\Psi_B$
are orthogonal in the infinitely heavy quark limit. The $1/m_Q$ 
corrections to the current also do not contribute at this 
kinematical point for the same reason.
The only nonzero contributions at $w=1$  come from corrections to the
Lagrangian~\footnote{There are no normalization conditions for 
these corrections
contrary to the decay to the ground state $D^{(*)}$ mesons, where the 
conservation of vector current requires their vanishing at
zero recoil \cite{luke}.} $\tilde\chi_1(w)$, $\tilde\chi_3(w)$ and 
$\chi_b(w)$. From Eqs.~(\ref{cff}) one can find for the form factors
contributing to the decay matrix elements at zero recoil 
\begin{eqnarray}\label{zeroc}
h_{+}(1)&=&\varepsilon_c\left[2\tilde\chi_1(1)+
12\tilde\chi_3(1)\right]+\varepsilon_b\chi_b(1),\cr
h_{A_1}(1)&=&\varepsilon_c\left[2\tilde\chi_1(1)-4\tilde\chi_3(1)\right]
+\varepsilon_b\chi_b(1).
\end{eqnarray}
Such nonvanishing contributions at zero recoil   
result from the first order $1 /m_Q$ corrections to the wave functions (see 
Eq.~(\ref{chib}) and the last terms in
Eqs.~(\ref{chi1}), (\ref{chi3})). Since the kinematically
allowed range for these decays is not broad ( $1\le w\le 
w_{\rm max}\approx 1.27$) the relative contribution to the decay rate of such 
small $1/m_Q$ corrections is substantially increased.  
Note that the terms 
$\varepsilon_Q(\bar\Lambda^{(n)}-\bar\Lambda)\xi^{(n)}(w)/(w-1)$ have
the same behaviour near $w=1$ as the leading order contribution, in contrast
to decays to the ground state $D^{(*)}$ mesons, 
where $1/m_Q$ corrections 
are suppressed with respect to the leading order contribution 
by the factor $(w-1)$ near this point (this result is known as Luke's 
theorem \cite{luke}). Since inclusion of first order heavy quark corrections to
$B$ decays to the ground state $D^{(*)}$ mesons results in approximately a 
10-20\% increase of decay rates \cite{fg,n}, one could expect that the 
influence of these corrections on decay rates to radially excited 
$D^{(*)}{'}$ mesons will be more essential. Our numerical analysis supports
these observations.

\begin{table}
\caption{Decay rates $\Gamma$ (in units  of $|V_{cb}/0.04|^2\times
10^{-15}$ GeV) 
and branching fractions BR (in \%) for  $B$ decays to radially
excited $D^{(*)}{'}$  mesons in the
infinitely heavy quark
limit and taking account of first  order $1/m_Q$ corrections. 
$\Sigma (B\to D^{(*)}{'}e\nu)$ 
represent the sum over both channels. 
$R'$ is a ratio
of branching fractions taking account of  $1/m_Q$ corrections to branching
fractions in
the infinitely heavy quark limit.  }
\label{tbr}
\begin{tabular}{cccccc}
\hline
   &\multicolumn{2}{c}{$m_Q\to\infty$}&\multicolumn{2}{c}{With $1/m_Q$}\\
Decay& $\Gamma$ & Br& $\Gamma$ & Br &$R'$ \\
\hline
$B\to D'e\nu$&0.53&0.12  & 0.92 & 0.22& 1.74\\
$B\to D^{*}{'}e\nu$&0.70&0.17 & 0.78 & 0.18& 1.11\\
$\Sigma (B\to D^{(*)}{'}e\nu)$ & 1.23& 0.29&1.70&0.40&1.37\\
\hline
\end{tabular}
\end{table} 

We can now calculate the decay  branching fractions by integrating double
differential  decay rates. Our results for decay rates
both in the infinitely heavy quark limit and taking account of the 
first order $1/m_Q$ corrections as well as their ratio
$$R'=\frac{{\rm Br}(B\to D^{(*)}{'}e\nu)_{{\rm with}\, 1/m_Q}}{{\rm Br}(B\to
D^{(*)}{'}e\nu)_{m_Q\to\infty}}$$
are presented in Table~\ref{tbr}. 
We find that both $1/m_Q$ corrections to decay
rates arising from corrections to HQET Lagrangian (\ref{chi1})--(\ref{chib}),
which do not vanish at zero recoil, and corrections to the current (\ref{xi3}),
 vanishing at zero recoil, give significant contributions. In
the case of $B\to D'e\nu$ decay both types of these corrections tend to
increase the decay rate leading to approximately a 75\% increase of the
$B\to D'e\nu$ decay rate. On the other hand, these corrections give opposite
contributions to the $B \to D^*{'}e\nu$ decay rate: the corrections to the
current give a negative contribution, while  corrections to the Lagrangian
give a positive one of approximately the same value. This interplay of
$1/m_Q$ corrections only slightly ($\approx 10\%$) increases the decay rate
with respect to the infinitely heavy quark limit. As a result the branching
fraction for $B\to D'e\nu$ decay exceeds the one for $B\to D^*{'}e\nu$
after inclusion of first order $1/m_Q$ corrections. In the infinitely heavy
quark mass limit we have for the ratio $Br(B\to D'e\nu)/Br(B\to D^*{'}e\nu)
=0.75$, while the account for $1/m_Q$ corrections results in the considerable 
increase of this ratio  to 1.22.

In Table~\ref{tbr} we also present the sum of the branching 
fractions over first
radially excited states. Inclusion of $1/m_Q$ corrections results in
approximately a 40\% increase of this sum. We see that our model predicts
that $ 0.40\%$ of $B$ meson decays go to the first radially excited $D$
meson states. If we add this value to our prediction for $B$ decays to
the first orbitally excited states $ 1.45\%$ \cite{orb}, we 
get the value of 1.85\%. This result means that approximately 2\% of
$B$ decays should go to higher charmed excitations.   

\section{Rare radiative $B\to K_2^*(1430)\gamma$ decay}

In the standard model $B\to K^{**}\gamma$ decays are governed
by the contribution
of the electromagnetic dipole operator $O_7$ to the effective Hamiltonian
which is obtained by integrating
out the top quark and $W$ boson and using the Wilson expansion \cite{lss}:
\begin{equation}\label{o7}
O_7=\frac{e}{16\pi ^2}\bar s\sigma ^{\mu \nu
}(m_bP_R+m_sP_L)bF_{\mu \nu }, \qquad P_{R,L}=(1 \pm \gamma _5)/2.
\end{equation}
The matrix elements of this operator between the initial $B$ meson state and
the final state of the orbitally
excited tensor $K_2^*$ meson have the following covariant decomposition
\begin{eqnarray}\label{ff1}
\langle K_2^*(p',\epsilon)|\bar s i k_\nu \sigma_{\mu \nu}b|B(p)\rangle
&=&i g_{+}(k^2)
\epsilon_{\mu \nu \lambda \sigma } \epsilon^{*\nu \beta}\frac{p_\beta}{M_B}
k^\lambda (p+p')^\sigma , \cr
\langle K_2^*(p',\epsilon)|\bar s i k_\nu \sigma_{\mu \nu}
\gamma_5b|B(p)\rangle & = &
g_{+}(k^2)\left(\epsilon_{\beta \gamma}^*\frac{p^\beta
p^\gamma }{M_B}(p+p')_\mu
-\epsilon_{\mu \beta}^*\frac{p^\beta }{M_B}(p^2-p'{}^2)   \right) \cr
& & +g_{-}(k^2)\left(\epsilon_{\beta \gamma}^*\frac{p^\beta
p^\gamma }{M_B}k_\mu
-\epsilon_{\mu \beta}^*\frac{p^\beta }{M_B}k^2\right) \cr
& &+h(k^2)((p^2-p'{}^2)k_\mu -(p+p')_\mu k^2)\epsilon_{\beta\gamma }^*
\frac{p^\beta p^\gamma  }{M_B^2 M_{K_2^*}},
\end{eqnarray}
where $\epsilon_{\mu\nu}$ is a polarization tensor of the final
tensor meson and $k= p -p'$ is the four momentum of the emitted
photon. The exclusive decay rate for the emission of a real
photon ($k^2=0$) is determined by the single form factor
$g_{+}(0)$ and is given by
\begin{equation}\label{drate}
\Gamma(B\to K_2^*\gamma)=
\frac{\alpha }{256\pi^4} G_F^2m_b^5|V_{tb}V_{ts}|^2
|C_7(m_b)|^2 g_{+}^2(0) \frac{M_B^2}{M_{K_2^*}^2}
\left(1-\frac{M_{K_2^*}^2}{M_B^2}
 \right)^5\left( 1+\frac{M_{K_2^*}^2}{M_B^2}\right),
\end{equation}
where $V_{ij}$ are the  Cabibbo-Kobayashi-Maskawa matrix elements and
$C_7(m_b)$ is the Wilson coefficient in front of the operator $O_7$.
It is convenient to consider the ratio of exclusive to inclusive
branching fractions, for
which we have
\begin{equation}\label{rk2}
R_{K_2^*}\equiv
\frac{BR(B\to K_2^*(1430)\gamma)}{BR(B\to X_s\gamma)}=
\frac18 g_{+}^2(0)\frac{M_B^2}{M_{K_2^*}^2}
\frac{\left(1-{M_{K_2^*}^2}/{M_B^2}
 \right)^5\left( 1+{M_{K_2^*}^2}/{M_B^2}\right)}{\left(1-{m_s^2}/{m_b^2}
 \right)^3\left( 1+{m_s^2}/{m_b^2}\right)}.
\end{equation}
The recent experimental value for the inclusive decay branching
fraction \cite{t}
$$BR(B\to X_s\gamma)=(3.15\pm 0.35\pm 0.41)\times 10^{-4}$$
is in a good agreement with theoretical calculations.

For the calculation of the decay matrix elements in our model we use
the same framework as in previous calculations of $B$ to excited $D$ 
decays. However, for a heavy-to-light transition we cannot expand
$\Gamma^{(2)}$ contribution  in inverse powers of the $s$ quark
mass. Instead we expand this contribution in inverse
powers of the large recoil momentum $|{\bm \Delta}|\sim
m_b/2$ of the final $K^{*}_2$ meson. 
The resulting expressions for the form factor $g_{+}(0)$ up to
the second order in $1/m_b$ can be found in \cite{rarexc}.

We can check the consistency of our expressions for $g_{+}(0)$ by taking
the formal limit of   $b$ and $s$ quark masses going to
infinity.~\footnote{As it was noted above  such limit is
justified only for the $b$ quark.} In this limit according to the
heavy quark effective theory \cite{vo2} the function
$\xi_F=2\sqrt{M_BM_{K_2^*}}g_{+}/(M_B+M_{K_2^*})$ should coincide
with the Isgur-Wise function $\tau$ for the semileptonic $B$
decay to the orbitally excited tensor $D$ meson, $B\to
D_2^*e\nu$. Such semileptonic decays have been considered by us
in Ref.~\cite{orb}. It is easy to verify that the equality of
$\xi_F$ and $ \tau$ is satisfied in our model if we also use the
expansion in $(w-1)/(w+1)$ ($w$ is a scalar product of
four-velocities of the initial and final mesons), which is small
for the $B\to D_2^*e\nu$ decay \cite{orb}. 
Calculating the ratio of  the form factor $g_+(0) $ in the
infinitely heavy $b$ and $s$ quark limit to the same form factor
in the leading order of  expansions in inverse powers of the
$b$ quark mass and large recoil momentum $|{\bm\Delta}|$
we find this ratio to be equal to $M_B/\sqrt{M_B^2+M_{K^*_2}^2}\approx
0.965$. The corresponding ratio for the form factor $F_1(0)$ of the exclusive
rare radiative $B$ decay to the vector $K^*$ meson \cite{gf} is equal to
$M_B/\sqrt{M_B^2+M_{K^*}^2}\approx 0.986$. Therefore we conclude
that the form factor ratios $g_+(0)/F_1(0)$ in the leading
order of these expansions differ by factor
$\sqrt{M_B^2+M_{K^*}^2}/\sqrt{M_B^2+M_{K^*_2}^2}\approx 0.98$.
This is the consequence of the relativistic dynamics leading to
the effective expansion in inverse powers of the $s$ quark energy
$\epsilon_s(p+\Delta)=\sqrt{({\bf p+\Delta})^2+m_s^2}$, which is
large in one case due to the large $s$ quark mass and in the
other one due to the large recoil momentum ${\bm \Delta}$.  As a
result both expansions give similar final expressions in the
leading order. Thus we can expect that the ratio $r$ from
(\ref{ratio}) in our calculations should be close to the one
found in the infinitely heavy $s$ quark limit \cite{vo}.

\begin{table}
\caption{ Our results in comparison with other theoretical
predictions and experimental data for branching fractions and their
ratios $R_{K^*}\equiv\frac{BR(B\to K^*\gamma)}{BR(B\to
X_s\gamma)}$,  $R_{K_2^*}\equiv\frac{BR(B\to
K_2^*\gamma)}{BR(B\to X_s\gamma)}$, $r\equiv\frac{BR(B\to
K_2^*\gamma )}{BR(B\to K^*\gamma)}$. Our values for the $B\to
K^*\gamma$ decay are taken from Ref.~\cite{gf}. } \label{tb}
\begin{tabular}{cccccc}
\hline
Value&our&Ref.~\cite{a}&Ref.~\cite{aom}
&Ref.~\cite{vo}&Exp.  \cite{cleo2}\\
\hline
$BR(B\to K^*\gamma)\times 10^5$& $4.5\pm1.5$
& 1.35 & $1.4 - 4.9$&$4.71\pm1.79$
&$4.55^{+0.72}_{-0.68}\pm0.34^a $ \\
 & & & & & $3.76^{+0.89}_{-0.83}\pm0.28^b$  \\
$R_{K^*}$ (\%) &$15\pm3$&4.5 &$3.5 - 12.2$ &$16.8\pm6.4$ &\\
$BR(B\to K_2^*\gamma)\times 10^5$&$1.7\pm0.6$
& 1.8 & $6.9 - 14.8$
& $1.73\pm0.80$& $1.66^{+0.59}_{-0.53}\pm0.13$ \\
$R_{K_2^*}$ (\%)
& $5.7\pm1.2$ & 6.0 & $17.3 -37.1$ & $6.2\pm2.9$ & \\
$r$
& $0.38 \pm 0.08$ & 1.3 & $3.0 - 4.9$ & $0.37\pm 0.10$
& $0.39^{+0.15}_{-0.13}$\\
\hline
\end{tabular}
\end{table}

The results of numerical calculations are given in Table~\ref{tb}. 
There we also show
our previous predictions for the $B\to K^*\gamma$ decay \cite{gf}.
Our results are confronted  with other theoretical calculations
\cite{a,aom,vo} and recent experimental data \cite{cleo2}. We
find a good agreement of our predictions for decay rates with the
experiment and estimates of Ref.~\cite{vo}. Other theoretical
calculations substantially disagree with data either for $B\to
K^*\gamma$ \cite{a} or for $B\to K^*_2\gamma$ \cite{aom} decay
rates. As a result our predictions and those of Ref.~\cite{vo}
for the ratio $r$ from (\ref{ratio}) are well consistent with
experiment, while the $r$ estimates of \cite{a}  and \cite{aom}
are several times larger than the experimental value (see
Table~\ref{tb}). As it was argued above, it is not accidental
that $r$ values in our and Ref.~\cite{vo} approaches are close.
The agreement of both predictions for branching fractions could be
explained by some specific cancellation of finite $s$ quark mass
effects and relativistic corrections which were neglected in
Ref.~\cite{vo}. We believe that our analysis is more consistent
and reliable. We do not use the ill-defined limit $m_s\to\infty$,
and our quark model consistently takes into account some
important relativistic effects, for example, the Lorentz
transformation of the wave function of the final $K^{*}_2$ meson
(see Eq.~(\ref{wig})). Such a transformation turns out to be very
important, especially for $B$ decays to orbitally excited mesons
\cite{orb}. The large value of the recoil momentum $|{\bm
\Delta}|\sim m_b/2$ makes relativistic effects to play a
significant role. On the other hand this fact simplifies our
analysis since it allows to make an expansion both in inverse
powers of the large $b$ quark mass and in the large recoil
momentum.

\acknowledgments
The authors express their gratitude to P. Ball, A. Golutvin, I. Narodetskii,
V. Savrin, and K. Ter-Martirosyan for discussions. 
D.E. acknowledges the support provided
to him by the Ministry of Education of Japan (Monbusho) for his
work at RCNP of Osaka University. Two of us (R.N.F and V.O.G.)
were supported in part by the {\it Deutsche
Forschungsgemeinschaft} under contract EB 139/2-1 and {\it
Russian Foundation for Fundamental Research} under Grant No.\
00-02-17768.

\end{document}